\documentstyle[aps,epsf,twocolumn]{revtex}

\begin{document}
\title{Transformation of dynamical fluctuation into 
coherent energy}
\author{Tsuyoshi Hondou}
\address{
Department of Physics, Tohoku University \\
Sendai 980-77, Japan
\\
(e-mail: hondou@cmpt01.phys.tohoku.ac.jp)
}
\maketitle

\begin{abstract}
 Studies of noise-induced motions are showing that 
coherent energy can be extracted
 from some kinds of noise in a periodic ratchet. 
 Recently, energetics of Langevin dynamics is formulated by
Sekimoto [J.Phys.Soc.Jpn, {\bf 66} 1234 (1997)], which can be applied 
 to 
ratchet systems
described by Fokker-Planck equation.
In this paper, we derive an energetics of ratchet systems
that can be applied to 
dynamical-noise-induced motion in a static potential.
Analytical efficiency of the energy transformation is derived
for the dynamical noise in an overdumping limit of the system.
 Comparison between analytical and numerical studies
is performed for chaotic noise.
\end{abstract}


 How efficient can noise be transformed into coherent
energy? 
 Recent advance of the studies of molecular 
motors\cite{A,Oosawa,ratchets,Hondou} wakes us up to the 
basic problem.
A famous trial to this problem can be found in the lecture note by
Feynman\cite{Feynman}, where he derived the efficiency of energy 
transformation in a system, what is called Feynman's ratchet, 
where multistability of the ratchet makes energy transformation
available.
 Recently, Sekimoto formulated energetics of 
Langevin equation\cite{Sekimoto}.
He applied it to the Feynman's ratchet and found an essential dissipation
mechanism that was lost in Feynman's discussion.
 
  The formulation of energetics of Langevin equation\cite{Sekimoto}
 is so generic that 
it may be analytically applicable to the system if it can be described
by Fokker-Planck 
equation\cite{Gardiner}, where an additive noise should be statistically simple.
 On the other hand, noise that breaks detailed balance of the applied system 
may often have statistically complex structure.
 A good example of such noise is chaos, where the fluctuation is generated
by dynamical process.
 However the effect of such dynamical noise on the 
multistable system
cannot be analyzed 
by Fokker-Planck equation for its un-negligible higher order time-correlation
function.
Thus, the energetics for these dynamical systems has not
been constructed yet, to my knowledge.

In this paper, we try to construct a theory that 
 describes the transformation 
of the dynamical noise into coherent energy by a periodic ratchet.
 First, we introduce a prototype system where the energy transformation
is available. 
We estimate in the next step the input
and the output energy and define the efficiency of the 
energy transformation. By introducing a dynamical noise that
is generated by a map, we derive an analytical expression of the efficiency
of energy transformation. 
For checking our method, we compare the analytical result
 with numerical one  using a chaotic map as a dynamical noise.
 With discussion for future 
studies, we conclude this paper.

Let us discuss a ratchet system where a particle moves in one dimensional 
periodic potential with 
resistance proportional to its velocity and an additive noise.
This system is described by  
a generalized Langevin equation,
\begin{equation}
m \frac{d^{2} x}{dt^{2}} = -\gamma \frac{dx}{dt} + \xi(t) 
- \frac{\partial V(x)}{ \partial x},
\end{equation}
where $m$, $\gamma$, $\xi$, are a mass of the particle, 
a resistance coefficient and an additive noise respectively. 
$V(x)$ is a periodic potential of which multistability makes it possible
 to transform noise into coherent energy.
We divide 
the noise term, $\xi(t)$, between  $\xi_{\mbox{T}}(t)$ and
$\xi_{\mbox{d}}(t)$:
$\xi(t) = \xi_{\mbox{T}}(t) + \xi_{\mbox{d}}(t)$, 
where $\langle \xi_{\mbox{T}}(t) \xi_{\mbox{T}} (s) 
\rangle = 2 \gamma k_{\mbox{B}} T
\delta(t-s)$.
The system does not move coherently if $\xi_{\mbox{d}} = 0$,
because the system then holds 
 "detailed balance". In such a case, the energy transformation is
not available,
which is the result of the second law of thermodynamics. Because the effect
of $\xi_{\mbox{d}}(t)$ causes finite drift of the system and therefore
energy transformation, 
 we consider zero temperature limit.
 To discuss the energy transformation, we add a mechanical load to the
periodic potential:
\begin{equation}
V(x) = V_{0}(x) + V_{l}(x),
\end{equation}
where $V_{0}(x)$ is a periodic potential and $V_{l}(x)$ is a potential
by which a load to the system is described.
We set $V_{l}(x) =l x$ for simplicity.
As a prototype, we adopt a piecewise linear
potential with parity symmetry
 as the periodic potential, $V_{0}(x)$:
 $V_{0}(x) = h-(h/L) |x( \, {
\rm mod}(2L))-L| $ for $ x \ge 0 $, $ V_{0}(-x) \equiv
V_{0}(x) $, where  $L$ is a half width of the period 
of the potential and $h$
is a height of the potential barrier.

To discuss the energetics of the present system,
we define  the input energy, $R$, from the noise 
to the system:
\begin{equation}
 R = \int_{x_{i}(t_{i})}^{x_{f}(t_{f})}  \xi(t) dx(t) \ .
\label{r}
\end{equation}
We also define a mechanical energy, $E_{l}$, that the system
acquires coherently in the fluctuating process:
\begin{equation}
 E_{l}  = V(x_{f},t_{f}) - V(x_{i},t_{i}) \ .
\end{equation}
 Then, the efficiency of the energy transformation, $\epsilon_{t}$,
between an initial state ($t_{i}$, $x_{i}$) and 
a final state ($t_{f}$, $x_{f}$)
is defined:
\begin{equation}
\epsilon_{t} = \frac{E_{l}}{R} \ .
\label{eps}
\end{equation}
The obtained 
coherent energy, $E_{l}$,
 is {\em globally} proportional 
 both to the strength of the load, $l$, and to 
the displacement of the system, $\Delta x$:
\begin{equation}
E_{l}  
= l \cdot \Delta x .
\label{135}
\end{equation}

In general, we have to solve
 Eq.(1)
to evaluate the efficiency of the energy transformation, Eq.(\ref{eps}).
 If the dominant noise was thermal, $\xi(t) \sim \xi_{\mbox{T}}$, 
Eq.(1) could be reduced to Fokker-Planck
equation, where average current could be obtained analytically.
 However,   when 
 the noise is dynamic, $\xi(t) \sim \xi_{d}(t)$, the reduction is not
 successful, because
the time-correlation function of the noise
 can not be truncated by second order.
Recently, 
it was found that an average current induced by chaotic
noise
 could be estimated analytically by focusing one's attention on the 
deterministic nature of the time series\cite{Hondou}.
Therefore we can proceed to discuss the energetics induced by dynamical 
noise. We introduce dynamical noise that is described by
a map:
\begin{equation}
\xi(t) = \Sigma_{i=-\infty}^{+\infty} F_{i} \delta(t-i),
\end{equation}
where $\{F_{i}\}$ is any dynamical time series produced by a map $f$:
\begin{equation}
 F_{i+1} = f (F_{i}).
\label{map}
\end{equation}

To evaluate the energy integral (Eq.\ref{r}), we 
consider first the effect of a single pulse at $t=t_{0}$
on $R$.
Energy input, $R_{0}$, by this pulse is:
\[ R_{0} \equiv \int F \delta(t-t_{0}) dx(t)
 = \int_{t_{0}-0}^{t_{0}+0} F \delta(t-t_{0}) 
\frac{dx}{dt} dt  \]
\begin{equation}
 = F \cdot (\frac{dx}{dt}|_{t=t_{0}-0}+\frac{dx}{dt}|_{t=t_{0}+0})/2.
\label{222}
\end{equation}
 By integrating Eq.(1) between $t=t_{0}-0$ and $t=t_{0}+0$, 
 we obtain through Eq.(\ref{222}):
\begin{equation}
 R_{0} = F \cdot (2 v_{0}+F/m)/2,
\end{equation}
where $v_{0} \equiv \frac{dx}{dt}|_{t=t_{0}-0}$.
 In case that the relaxation time of the velocity, 
$m/\gamma$, is smaller than
the unit interval of the pulse, $m/\gamma \ll 1$,
 the effect of the velocity, $v_{0}$, on  $R_{0}$ is negligible. 
As we will consider
the overdumping motion, we will neglect the effect of $v_{0}$.
Average input energy per unit time
is obtained using 
an invariant density of the map:
\begin{equation} 
 \langle R \rangle \simeq \langle \frac{F_{i}^{2}}{2m} 
\rangle = \frac{1}{2m} 
\int F^{2} \rho (F) dF,
\label{110}
\end{equation}
where $\rho (F)$ is the invariant density of the map.
The average displacement, $\langle \Delta x \rangle$, in the unit 
time is: 
\begin{equation}
 \langle \Delta x \rangle = P_{tr} \cdot 2L ,
\label{111}
\end{equation} 
where $P_{tr}$ is a probability that a particle crosses a peak of the 
periodic potential per unit time. 
We call $P_{tr}$ as "barrier crossing probability" hereafter
 that is obtained in a theory of chaotic noise\cite{PRE}.
Inserting Eq.(\ref{111}) into Eq.(\ref{135}),
we obtain the average coherent energy, $\langle E_{l} \rangle$,
that the system acquires in a unit time:
\begin{equation}
\langle E_{l} \rangle = P_{tr} \cdot 2L \cdot l  .
\label{112}
\end{equation}
Inserting Eqs.(\ref{110}) and (\ref{112}) into Eq.(\ref{eps}), we obtain
 the average efficiency of the energy transformation:
\begin{equation}
 \langle \epsilon_{t} \rangle = \frac{ l \cdot P_{tr} \cdot 2L}{\frac{1}{2m} \int 
                F^{2} \rho(F) dF}.
\label{formula}
\end{equation}

In the derivation of the efficiency, we have 
introduced some approximations.
Thus, we will compare this analytical
results with numerical ones.
For demonstration, we use tent map chaos\cite{Schuster} as a dynamical noise.
We set:
\begin{equation}
F_{i+1}=1/2 - 2 | F_{i} |.
\end{equation}
The tent map has a constant invariant density and $\delta$-correlation
function, of which the simple statistical character is useful
for an analysis.
In Fig.1, we show the typical dependence of the efficiency of
energy transformation on the strength of the load.
 The numerical result shows that the efficiency 
monotonically increases up to $ l \sim 10^{-2}$; and after the peak, 
the efficiency decreases suddenly. The present theory sufficiently
predicts the increase of the efficiency up to the peak. The numerical
result shows that decrease of the efficiency starts faster than that
of theory. This disagreement is attributed mainly to the disagreement 
of the barrier crossing probability, $P_{tr}$, that
is estimated by a previous theory\cite{PRE}: Before
the efficiency starts saturation, the system crosses the potential
barrier unidirectionally.
In this regime, the theory of chaotic noise\cite{PRE} precisely
predicts the barrier crossing probability, $P_{tr}$.
However, when the saturation begins, the system crosses it in both
directions. In this regime, the theory of chaotic noise does not
precisely predicts the barrier crossing probability; the disagreement 
of the efficiency emerges in this regime.

In case that the potential barrier is $60\%$ less than that of Fig.1, 
the efficiency
of energy transformation (Fig.2) is higher than that of Fig.1.
 Although the order
of energy transformation of Fig.2 differs from Fig.1, the functional
forms of the two Figures are almost the same.
The difference of the  maximum efficiencies of energy transformation
between Fig.1 and Fig.2 is attributed to the difference of
the barrier crossing probabilities.
\begin{figure}[h]
\epsfile{file=fig1c.EPSF,scale=0.35}
\caption{Efficiency of energy transformation of a system
where $m = 0.1$, $\gamma = 1.0$,
$L=5.0$ and $h=0.5$.
}
\end{figure}
\begin{figure}[h]
\epsfile{file=fig2c.EPSF,scale=0.35}
\caption{Efficiency of energy transformation of a system 
where $m = 0.1$, $\gamma = 1.0$,
$L=2.0$ and $h=0.2$.
}
\end{figure}

 In this paper, we have derived an energetics of
a ratchet system that transforms dynamical noise to coherent energy.
The formula of the 
efficiency of energy transformation itself would be certainly 
robust as far as overdumping condition, $m/\gamma \ll 1$ holds: 
the formula (Eq.\ref{formula}) is not restricted to special shape
of the potential.
The
precision of the efficiency much depends on the precision of the
barrier crossing probability, $P_{tr}$.

 From the viewpoint of dynamical systems, one can imagine that
 the efficiency
should reflect the complexity of the noise.  The complexity can be
estimated by, for example, the Kolmogorov Entropy\cite{Schuster2}.
We conjecture that high efficiency would emerge when the Kolmogorov
Entropy of the dynamical noise is low. If the functional form (Eq.\ref{map})
of the noise is restricted, we can find an realization of this 
conjecture\cite{Hondou4}. We are now at the starting point of this analysis.

 Real protein motors is known to operate at surprisingly high efficiency of 
energy transformation\cite{Woledge}.
 However, none of the ratchet models have not successfully
 explained this high efficiency yet, because energetics had not been 
derived in ratchet models until Ref.\cite{Sekimoto}
except for Ref.\cite{Oosawa}, to my knowledge.
 We have not studied to optimize the efficiency
by varying the shape of the potential and parameters of the system.
Further study of the optimal efficiency of ratchet models will 
 bring us a new insight as to the biologically possible mechanism of 
 protein motors. 

 We gratefully acknowledge K. Sekimoto, Y. Sawada, S. Nasuno, S. Sasa,
 T. Tsuzuki for helpful comments
and discussions and also acknowledge T. Fujieda and F. Takagi 
for critical reading of 
the paper. 
  Numerical calculation was partly performed by
the workstations of YITP, Kyoto University. This work is supported in part
by the Japanese Grant-in-Aid for Science Research Fund from the Ministry
of Education, Science and Culture (No. 09740301).


\begin{thebibliography}{9}
\bibitem{A}
Early works where the idea of noise-induced transport was discussed are
found in
M. B\"uttiker, Z. Phys. {\bf B 68}, 161 (1987);
R. Landauer, J. Stat. Phys. {\bf 53} 233 (1988).
\bibitem{Oosawa}
 Efficiency of energy transformation was first discussed  
 in a ratchet model of 
molecular motor: R. D. Vale and F. Oosawa, Adv. Biophys. {\bf 26}, 97 (1990).
 Because the efficiency derived in this paper was based on the incorrect 
calculation 
by Feynman\cite{Feynman}, the efficiency was estimated much higher than 
that of 
Ref.\cite{Sekimoto}. 
\bibitem{ratchets}
For examples: 
M. O. Magnasco,  Phys. Rev. Lett. {\bf 71}, 1477 (1993);
J. Rousselet, L. Salom\'e, A. Ajdari, and J. Prost, 
Nature {\bf 370},
446 (1994);
A. Ajdari, D. Muramel, L. Peliti, and J. Prost, J. Phys. I (France)
{\bf 4}, 1551 (1994);
C. R. Doering, W. Horsthemke, and J. Riordan, Phys. Rev. Lett. {\bf 72},
 2984 (1994).
R. D. Astumian and M. Bier, Phys. Rev. Lett. {\bf 72}, 1776 (1994);
 L. Faucheux, L. Bourdieu, P. Kaplan, and A. Libchaber,
 Phys. Rev. Lett. {\bf 74},
 1504 (1995).
\bibitem{Hondou}
T. Hondou,
  J. Phys. Soc. Jpn. {\bf 63}, 2014 (1994);
 T. Hondou and Y. Sawada,  Phys. Rev. Lett. {\bf 75}, 3269 (1995); {\em ibid.}
{\bf 76}, 1005 (1996).
\bibitem{Feynman}
R. P. Feynman, R. B. Leighton, and M. Sands, {\em The Feynman Lectures
in Physics} (Addison-Wesley, Massachussets, 1966).
\bibitem{Sekimoto}
 K. Sekimoto, J. Phys. Soc. Jpn. {\bf 66}, 1234 (1997).
\bibitem{Gardiner}
 See, for example, C. W. Gardiner, 
{\em Handbook of Stochastic Methods 2nd ed.}
(Springer-Verlag, Berlin, 1990).
\bibitem{PRE}
 Formulae of barrier crossing probability, $P_{tr}$, for a generalized map
 are given  in
 T. Hondou and Y. Sawada, Phys. Rev. {\bf E54}, 3149 (1996). 
\bibitem{Schuster}
 S. Grossmann and H. Fujisaka, Phys. Rev. {\bf A26}, 1779 (1982).
\bibitem{Schuster2}
 See, for example, H. G. Schuster,  {\em Deterministic chaos}, (Physik-Verlag,
Berlin, 1984).
\bibitem{Hondou4}
 T. Hondou, (unpublished).
\bibitem{Woledge}
 For example, R. C. Woledge, N. A. Curtin and E. Homsher, 
{\em Energetic Aspect of Muscle Contraction} (Academic Press, 
New York, 1985). 
\end{thebibliography}
\end{document}